\documentclass[aps,twocolumn,showpacs, prb]{revtex4}
\usepackage{graphicx}
\usepackage{bm}
\usepackage{amsmath}

\providecommand{\ket}[1]{\lvert #1 \rangle}

\providecommand{\be}{\begin{equation}}
\providecommand{\ee}{\end{equation}}
\providecommand{\ba}{\begin{eqnarray}}
\providecommand{\ea}{\end{eqnarray}}
\begin{document}

\title{Direct measurement of the biphoton Wigner function through two-photon interference}
 
\author{T. Douce$^1$, A. Eckstein$^1$, S. P. Walborn$^2$, A. Z. Khoury$^3$, S. Ducci$^1$, A. Keller$^4$, T. Coudreau$^1$ and P. Milman$^1$}

\affiliation{$^{1}$Laboratoire Mat\'eriaux et Ph\'enom\`enes Quantiques, Universit\'e Paris Diderot, CNRS UMR 7162, 75013, Paris, France}
\affiliation{$^{2}$ Instituto de F\'{\i}sica, Universidade Federal do Rio de Janeiro. Caixa Postal 68528, 21941-972 Rio de Janeiro, RJ, Brazil}
\affiliation{$^{3}$ Instituto de F\'isica, Universidade Federal Fluminense, 24210-340 Niter\'oi-RJ, Brasil}
\affiliation{$^{4}$Univ. Paris-Sud 11, Institut de Sciences Mol\'eculaires d'Orsay (CNRS), B\^{a}timent 350--Campus d'Orsay, 91405 Orsay Cedex, France}


\vskip2pc 
 
\maketitle
{\bf The Hong-Ou-Mandel (HOM) \cite{HOM} experiment was a benchmark in quantum optics, evidencing the quantum nature of the photon. In order to go deeper, and obtain the complete information about the quantum state of a system, for instance, composed by photons, the direct measurement or reconstruction of the Wigner function\cite{WIGNER}  or other quasi--probability distribution in phase space is necessary. In the present paper, we show that a simple modification in the well-known HOM experiment provides the direct measurement of the Wigner function.  We apply our results to a widely used quantum optics system, consisting of the biphoton generated in  the parametric down conversion process. In this approach, a negative value of the Wigner function is a sufficient condition for  non-gaussian entanglement between  two photons. In the general case, the Wigner function provides all the required information to infer entanglement using well known necessary and sufficient criteria \cite{SIMON}. We analyze our results using two examples of parametric down conversion processes taken from recent experiments \cite{ANDREAS, STEVE2}. The present work offers a new vision of the HOM experiment that further develops its possibilities to realize fundamental tests of quantum mechanics involving decoherence and entanglement using simple optical set-ups. 
}
\par
Entangled photon pairs play undoubtedly a central role in quantum information processing and quantum communication. Photons are the most efficient quantum information carriers, not only for their intrinsic propagation speed, but also for the variety of degrees of freedom they possess, both discrete and continuous.  Some examples of quantum information protocols that have been realized with photons are teleportation \cite{Bouwmeester97}, quantum key distribution \cite{Gisin02}, one--way quantum computing  \cite{Walther05} and quantum repeaters \cite{REPEATERS}. Moreover, entangled photon pairs enable the realization of fundamental tests of quantum mechanics, as Bell type inequalities \cite{BELL}, since the no--signalling condition is relatively easily fulfilled. 

Spontaneous parametric down conversion (SPDC) is the most widely used process to generate entanglement in different (independent) degrees of freedom of a photon pair. Detecting, characterizing and manipulating this entanglement is a key issue for quantum information applications. This problem is fundamentally different if one is dealing with discrete degrees of freedom (e.g. polarization), or with continuous ones (e.g. spatial or spectral). While for two qubit states in the discrete case and for gaussian states in the continuous one, necessary and sufficient conditions exist for entanglement detection, for higher dimensions or more general configurations,  solutions are subspace dependent \cite{VOGEL}. However, using high dimensional systems and non Gaussian states leads to a number of important and interesting applications, such as entanglement distillation \cite{TAKAHASHI09}, quantum computation \cite{COMP} and high precision measurement \cite{DALVIT}. For these reasons, understanding and classifying such states is a matter of importance and  fundamental interest. 
\par
Photons produced by SPDC can be highly non-separable because the characteristics of the pump beam and of the nonlinear medium are transferred to global degrees of freedom of the photon pair. This transfer also occurs in the strong field regime where
the quadratures of the down converted fields are entangled
and could present non Gaussian behaviour for sufficiently
high nonlinear coupling \cite{kaled}. In the photon pair regime, we often speak of \emph{biphoton} states. For instance, using a continuous wave (cw) pump and considering degenerate, monochromatic and polarized fields, the two-photon state can be written as 
\cite{REVIEW,STEVE2}
\begin{equation}\label{sinc}
 \ket{\psi}\!=\!\iint \!\! F_+ ({\bf p}_1+{\bf p}_2) F_-\left ({\bf p}_1-{\bf p}_2\right )\!\ket{{\bf p_1},{\bf p_2}}{\rm d}{\bf p_1}{\rm d}{\bf p_2},
 \end{equation}
where $F_+$ is the normalized momentum distribution of the pump beam, $F_-$ is the phase matching function and ${\bf p_i}$ the Transverse Momentum (TM) vector of the $i$-th photon \cite{REVIEW}.  Eq. \eqref{sinc} can indeed be obtained for several types of continuous variables (namely spatial or frequency coordinates) and in a wide range of experimental setups, as we will show below. We denote $\bf{p}_\pm=\bf{p_1}\pm\bf{p_2}$ and $\bf{q}_\pm=\bf{q_1}\pm\bf{q_2}$ (the sum and differences of position coordinates). The correlations of the biphoton are determined by the functions $F_{\pm}(\bf{p}_\pm)$ and their Fourier transforms $\mathcal{F}_\pm({\bf q}_{\pm})$, which describe the photons in the transverse position coordinate.    
To gain information about the entanglement in state \eqref{sinc}, measurement of the coincidence distributions in at least Fourier conjugate planes is required.  Furthermore, to gain total information about the quantum state through tomography requires measurement of correlations along additional directions in the phase space of the transverse spatial variables \cite{BANAZSEK}. This could be done, in principle, by generalizing the method demonstrated in \cite{BANAZSEK2} where the Wigner function of a single photon was directly measured using a Sagnac interferometer. This method can, in principle,  lead to the measurement of the Wigner function for photons prepared in an arbitrary state, but it demands the stabilization of independent  interferometers. The method to measure the biphoton's Wigner function presented here, as will be seen in the following has a much higher stability in spite of the symmetry conditions required for the wave-function. Moreover, it is based on a currently used technique to probe the quantum statistical properties of bosons and fermions.  It is obvious from Eq.~\eqref{sinc} that all of the information about correlations between photons 1 and 2 can be obtained through measurements on the sum and difference variables, ${\bf p}_{\pm}$. 

In the present paper, we show that the biphoton Wigner function  can be measured directly using an adaptation of the Hong-Ou-Mandel (HOM) interferometer \cite{HOM}.  In the following, we detail how to detect the biphoton Wigner distribution in the whole phase space and consequently, obtain a full characterization of the two-photon quantum state, by using readily available linear optics elements.  Furthermore, we show that HOM interference can be used as an entanglement witness for non-Gaussian entanglement. Negative values of the biphoton Wigner function is a sufficient condition to prove the presence of non-Gaussian entanglement in the system. 
\par
\begin{figure}[h]
\centerline{\includegraphics[width=.75\columnwidth]{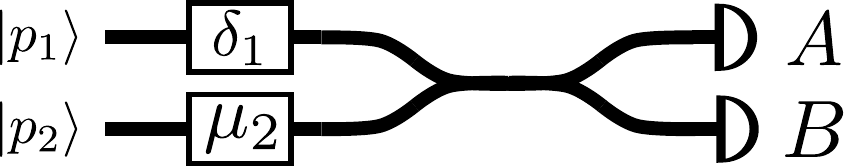}}
\caption{Scheme of the HOM-type interferometer. Devices represented by boxes in each arm displace the continuous variable degree of freedom in conjugate spaces, so that the Wigner function can be measured in all the points of phase space \label{fig1}} 
\end{figure}

Let us first recall the principles of the HOM interferometer, shown in Fig \ref{fig1}. Each photon of a pair created from SPDC is sent through one of the two arms of an interferometer. They are then recombined in a $50/50$ beam splitter (BS) and detected by detectors A and/or B. When the photons are indistinguishable and reach the BS simultaneously, they bunch and  follow the same path. Coincidence detections in detectors in A and B are thus less likely, and the so-called ``Hong-Ou-Mandel dip" is observed \cite{HOM}.  

For notational simplicity, we write the state of a continuous variable degree of freedom of the photons as $\ket{\Psi}=\iint F(p_1,p_2) \ket{p_1,p_2}{\rm d}p_1 {\rm d}p_2$, where $p_i$ are variables associated to the $i$-th photon ($i=1,2$) that propagates in the $i$-th arm of the interferometer. We impose that all other degrees of freedom are separable from the considered one, which can be guaranteed using filters to make local projections. In order to better illustrate the main idea, we will first take $p_i$ as being the TM of the $i$-th photon produced in the SPDC process, and consider only one spatial dimension~\cite{REVIEW}. Examples of applications to other continuous degrees of freedom and extension to two dimensions will be given below.
\smallskip
\section{Results}

In the present version of the HOM experiment, we suppose the interferometer is calibrated and both photons reach the BS simultaneously. We then add a position translation $2\delta$ to photon labeled $2$ and a momentum translation $\mu$  to photon labeled $1$. These operations are currently done using linear optical elements \cite{STEVE,SI}. After both position and momentum displacements, the biphoton state, impinging in the BS is:
\begin{equation}
\ket{\Psi}\rightarrow \ket{\Psi_{\mu,\delta}}=\iint  F(p_1,p_2) e^{-2i p_2\delta} \ket{p_1+\mu,p_2}{\rm d}p_1 {\rm d}p_2.
\end{equation}
\begin{widetext}
After the BS, the  two-photon state  is given by
\begin{equation}
\ket{\Psi_{BS}}=1/2  \iint \mathrm d p_1\mathrm d p _2 \, F(p_1,p_2) e^{-2 ip_2 \delta } (\ket{p_1- \mu}_A\ket{p_2}_B-\ket{p_1 - \mu}_B\ket{p_2}_A+\ket{p_1- \mu}_B\ket{p_2}_B-\ket{p_1- \mu}_A\ket{p_2}_A).
\end{equation} 	 
We will now focus on the coincidence detections only, {\it i.e.}, consider only states corresponding to two photons exiting in different paths, A and B. The coincidence probability $I(\mu,\delta)$ thus reads
\begin{equation}\label{HOM}
I(\mu,\delta)\!=\!\frac{1}{2}-\frac{1}{2} {\rm Re}\!\left[\iint F(p_2,p_1)F^\ast (p_1+\mu,p_2-\mu)e^{-2 i\, (p_1 -p_2) \,\delta}{\rm d}p_1 {\rm d}p_2\right]\!\!.
\end{equation} 
\end{widetext}
Let us already note that separability of the biphoton wavefunction $F(p_1,p_2)=f_1(p_1)f_2(p_2)$ implies  $I( \mu,\delta)\!\leq\!1/2$.
This applies to two-photon mixed states as well, since they can be constructed as a convex sum of separable pure states \cite{SI}. Thus $I(\mu,\delta) \leq 1/2$ is an entanglement witness for general two-photon state of the TM. A similar result was obtained for other degrees of freedom in Refs.~\cite{VANENK, ANDREAS, PIRES}.

Let us now turn to the main result of our paper. For the sake of simplicity, we will independently discuss each transverse axis, $x$, parallel to the incidence plane, and $y$, orthogonal to the incidence plane.  This is necessary since there is a fundamental difference between $x$ and $y$ axes under reflection upon a vertical mirror since $p_x\rightarrow - p_x$ and $p_y\rightarrow p_y$, which has interesting consequences on our results, as will be seen below \cite{COMMENT}. 

We start by assuming, as in Eq. (\ref{sinc}), $F(p_{1,i},p_{2,i})=F_-(p_{-,i})F_+(p_{+,i})$, where $i=x,y$. This assumption is verified in most experiments with SPDC and is commonly used when studying entanglement in this process  \cite{REVIEW,LAW}.  From now on, in order to simplify the notation, we will index functions instead of variables so that $F_+(p_{+,i}) \equiv F_{i+}(p_+)$, for instance. 

The integral in Eq. (\ref{HOM}) reads differently for $x$ or $y$ coordinates \cite{COMMENT}.  
For the $y$ axis, we have
\begin{eqnarray}\label{omegay}
&& \int \left\vert F_{y+}(\mu + p_+) \right\vert^2 \mathrm{d}p_+ \times \\
&& \int F_{y-}(\mu + p_-)F_{y-}^{*}(\mu - p_-) e^{-2 i  p_-\delta} \mathrm{d}p_-,  \nonumber 
\end{eqnarray}
while for the $x$ axis we have
\begin{eqnarray}\label{omegax}
&& \int | F_{x-}(\mu + p_-)|^2 \mathrm{d}p_- \times \\
&&\int F_{x+}(\mu + p_+) F_{x+}^{*}(\mu - p_+)e^{-2 i p_+\delta}\mathrm{d}p_+.  \nonumber 
\end{eqnarray}
First, let us notice that normalization can be chosen so  that the first integrals in Eqs. \eqref{omegay} and \eqref{omegax} are unity (integrals in $p_+$ and $p_-$, respectively).
Eq.~(\ref{omegay}) becomes
\begin{equation}\label{wigner1}
 \int F_{y-}(\mu + p_-)F_{y-}^{*}(\mu - p_-) e^{-2 i  p_-\delta} \mathrm{d}p_-= \pi W_{y-}(\mu,\delta),
\end{equation}
while Eq. (\ref{omegax}) becomes:
\begin{equation}\label{wigner1b}
\int F_{x+}(\mu + p_+) F_{x+}^{*}(\mu - p_+)e^{-2 i p_+ \delta}\mathrm{d}p_+= \pi W_{x+}(\mu,\delta).
\end{equation}
where $W_{y-}(\mu,\delta)$ and $W_{x+}(\mu,\delta)$  are, by definition, the Wigner functions at point $(\mu,\delta)$~\cite{WIGNER} associated to wave functions $F_{y-}$ or $F_{x+}$, respectively.

Thus, the coincidence probability in this adapted version of the HOM experiment reveals the Wigner function at phase space point $(\mu,\delta)$:
\begin{equation}\label{cprob}
I(\mu,\delta) = \frac{1}{2} -  \frac{\pi}{2} W_{j}(\mu,\delta),
\end{equation}
where $j=x+$ or $j=y-$. We stress that, in the case where space components are not separable and/or the wave-function is not separable in the  ``+" and ``-"  coordinates, our main result still holds: the proposed adaptation of the HOM experiment leads to the Wigner function of the bi--photon. Except that in this case, we experimentally access specific regions of the phase space \cite{SI}. 
Also, it is a straightforward calculation to show that the Wigner function of a non-pure state can also be directly measured using the HOM set-up described above \cite{SI}. 

From the Wigner function one can infer all the necessary information about the state, and in particular, entanglement for Gaussian and non-Gaussian states. The witness defined by $I(\mu,\delta) \leq 1/2$ allows to detect non-Gaussian states since they may have negative values of the Wigner function. Though the witness itself does not detect gaussian entanglement, we can nevertheless use the Wigner function to test Gaussian entanglement using other criteria \cite{SIMON}.  These facts, added to the one that no assumption is being made on the width of the distribution, are clear advantages of the present method over discretization based techniques for detecting entanglement in continuous variable systems \cite{LAW}. 

We note that the Wigner function appearing in (\ref{wigner1}) is a single-party Wigner function referring to the sum or difference coordinates of the biphoton. This is a direct consequence of the form of state \eqref{sinc} and momentum conservation.   
\par
We have shown that the reflexion asymmetry of the TM correlates the measurement  of the entanglement properties of $F_+$ or $F_-$ to orthogonal traverse directions (Eqs. (\ref{wigner1}) and (\ref{wigner1b})). However, one can measure these functions in {\it either} axis, since they can be controllably interchanged, for instance by adding a Dove prism orientated at $45^\circ$ in both arms of the HOM interferometer that rotates the fields by $90^\circ$. 
\par
Up to now, we have independently considered each degree of freedom of the biphoton, but the spatial variables are inherently two-dimensional (2D), as exemplified by Eq.~\eqref{sinc}. This leads to a four-dimensional Wigner function, instead of a two-dimensional one. Using the obtained results and considering the reflection properties of the BS, it is straightforward to show that the four-dimensional Wigner function returns information about $F_+$ in the $x$ direction, and about $F_-$ in the $y$ direction. If we consider, for instance,  a two-photon state of the form \eqref{sinc}, in  the approximation where both transverse coordinates are separable, the application of our results to  the  two dimensions simultaneously, gives \cite{SI}
\begin{equation}\label{eq:2d}
I(\mu_x,\delta_x; \mu_y, \delta_y) = \frac{1}{2} - \frac{\pi^2}{2} W_{x+}(\mu_x,\delta_x) W_{y-}(\mu_y, \delta_y). 
\end{equation}
Using Dove prisms can lead to similar expressions as (\ref{eq:2d}) involving orthogonal coordinates, as mentioned above. It is also worth mentioning that even if transverse coordinates are not separable, $I(\mu_x,\delta_x; \mu_y, \delta_y)$ reveals the Wigner function of each transverse coordinate. In this case,  we are measuring the Wigner function of non pure states, as detailed in \cite{SI}. 

\begin{figure}[h]
$\begin{array}{cc}
(a)\includegraphics[width=3.5cm]{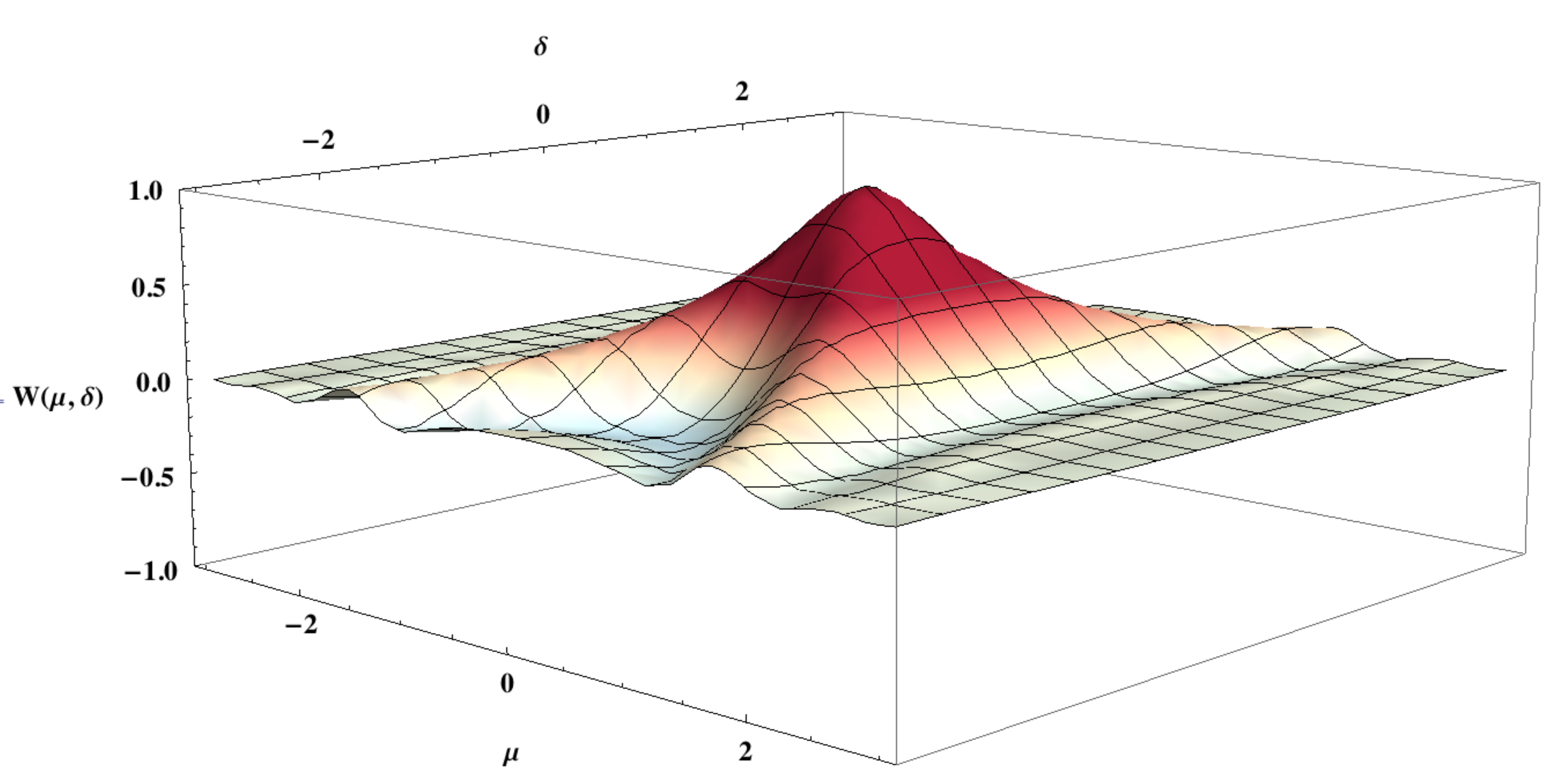}&
(b)\includegraphics[width=3.5cm]{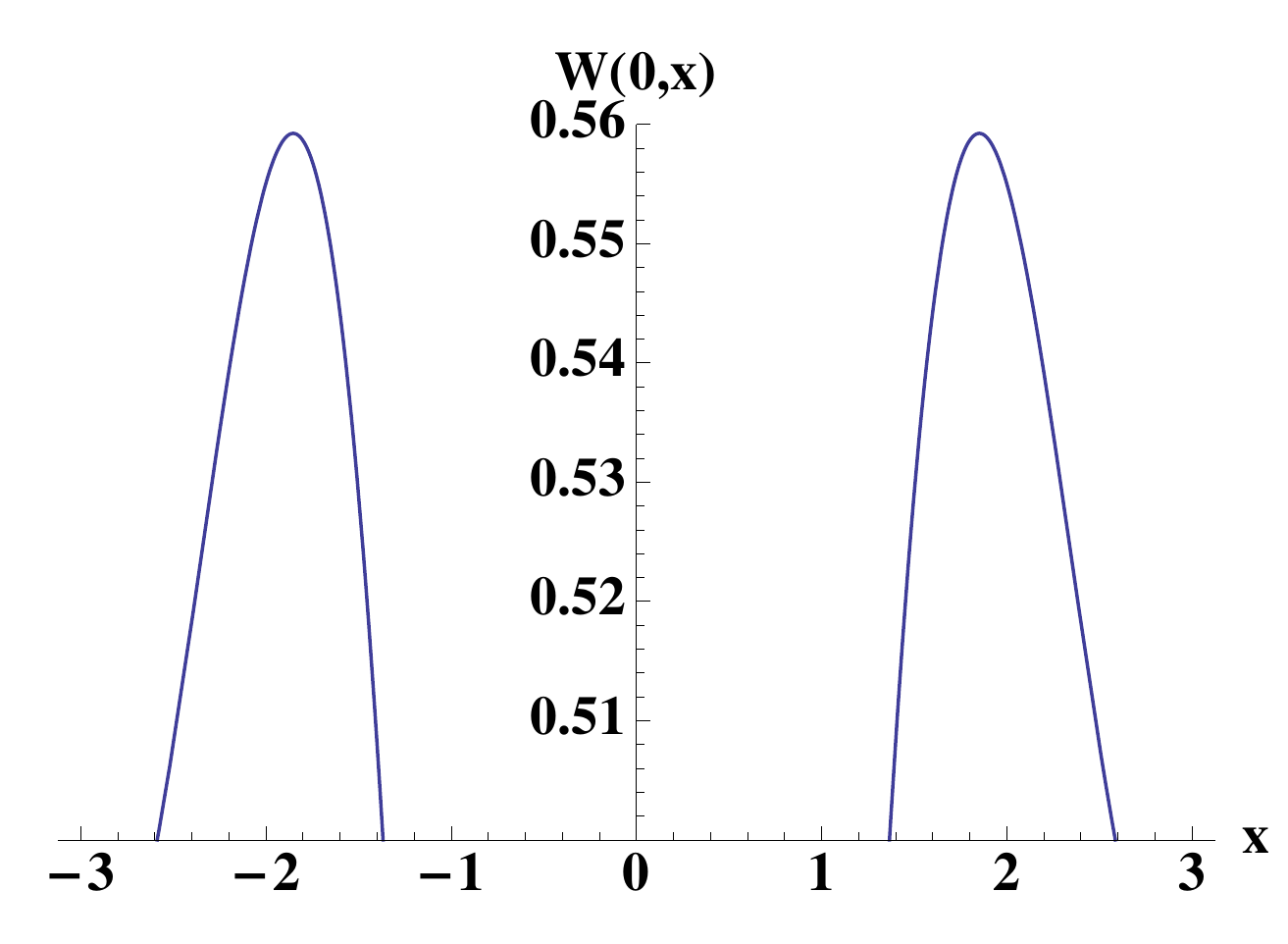} \\ 
(c)\includegraphics[width=3.5cm]{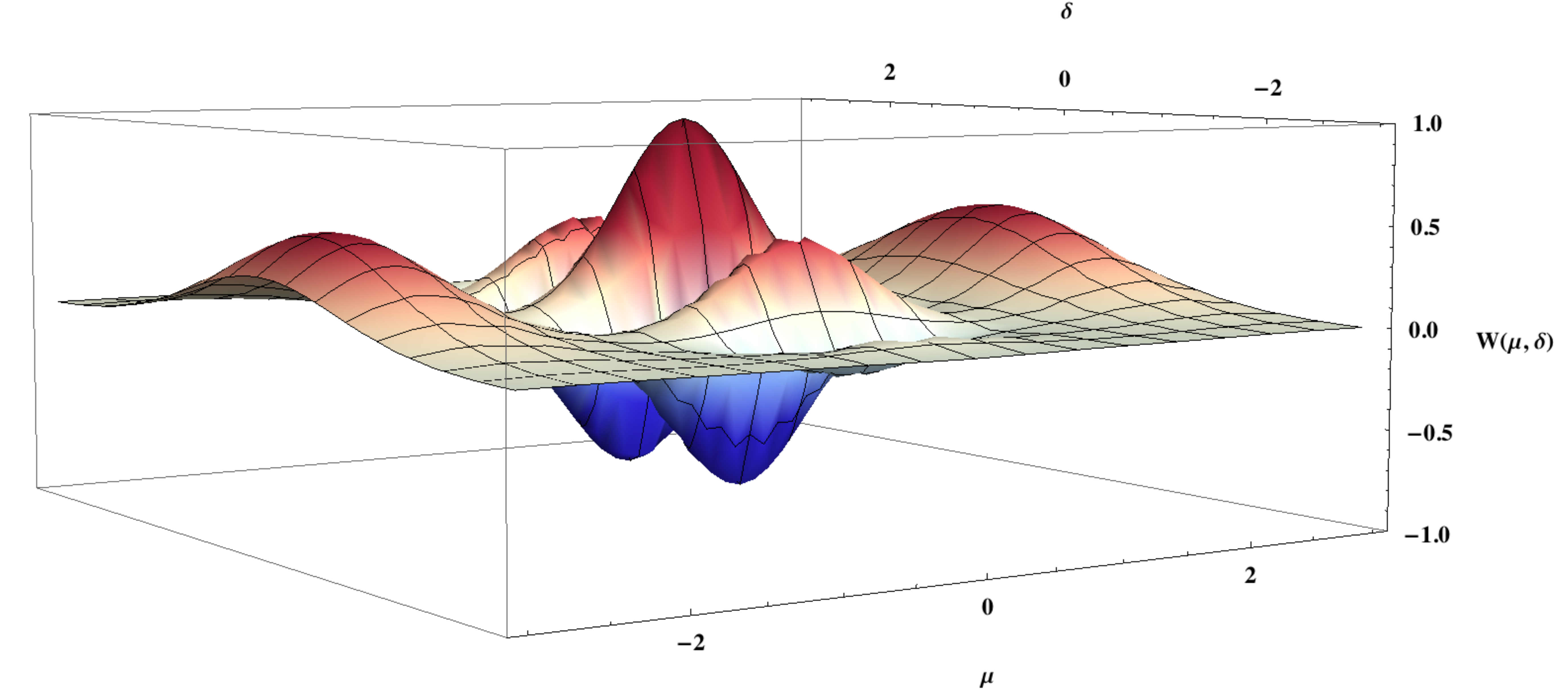} &
(d)\includegraphics[width=3.5cm]{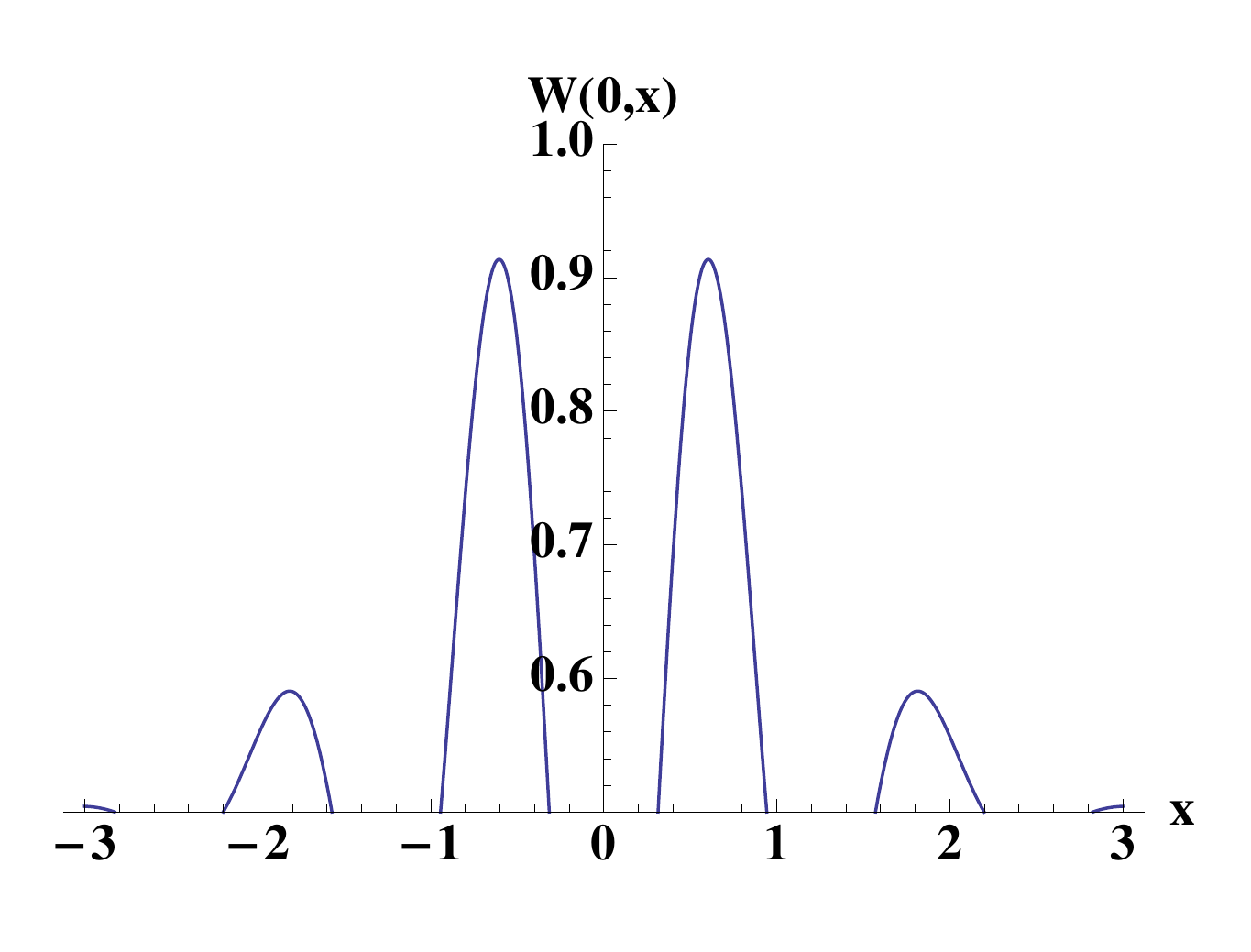}\\
(e)\includegraphics[width=3.5cm]{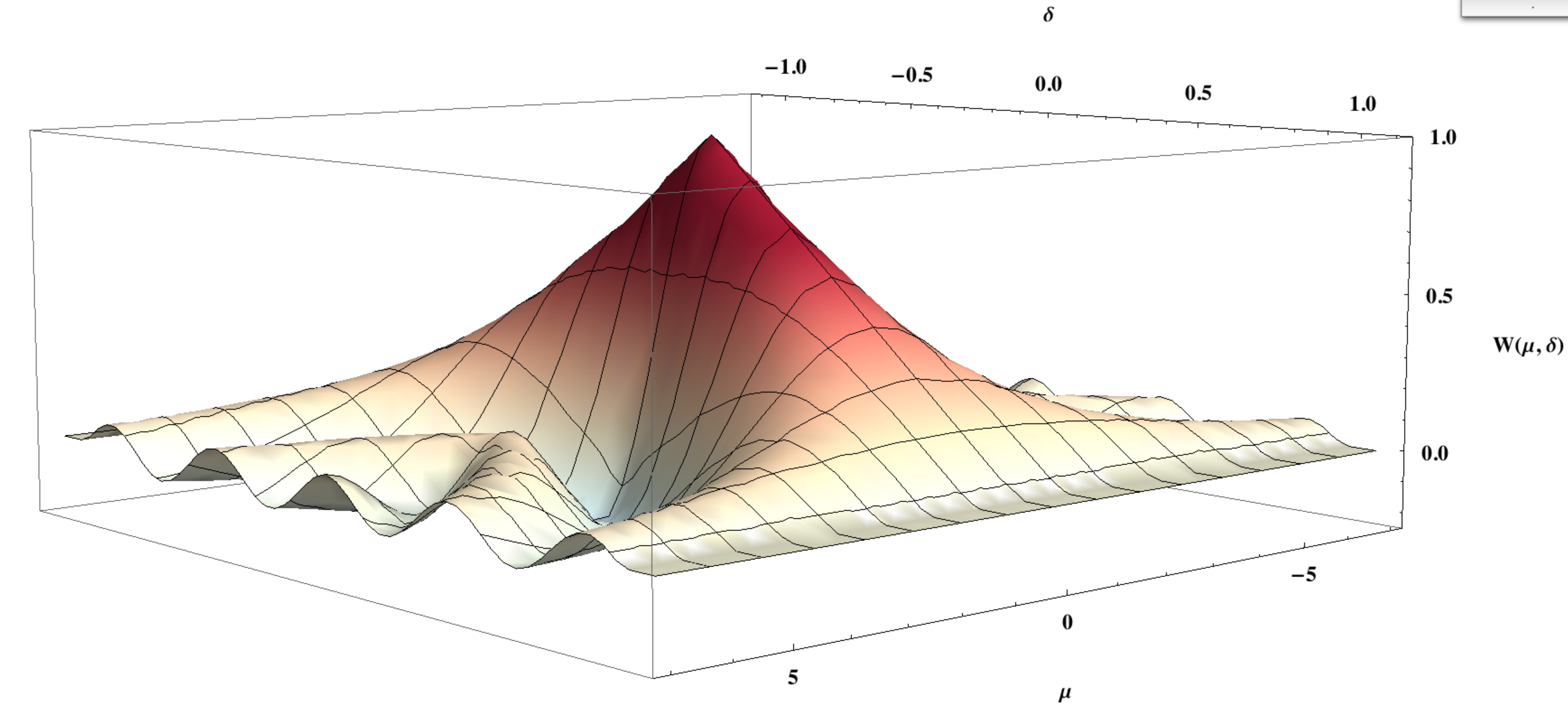} &
(f)\includegraphics[width=3.5cm]{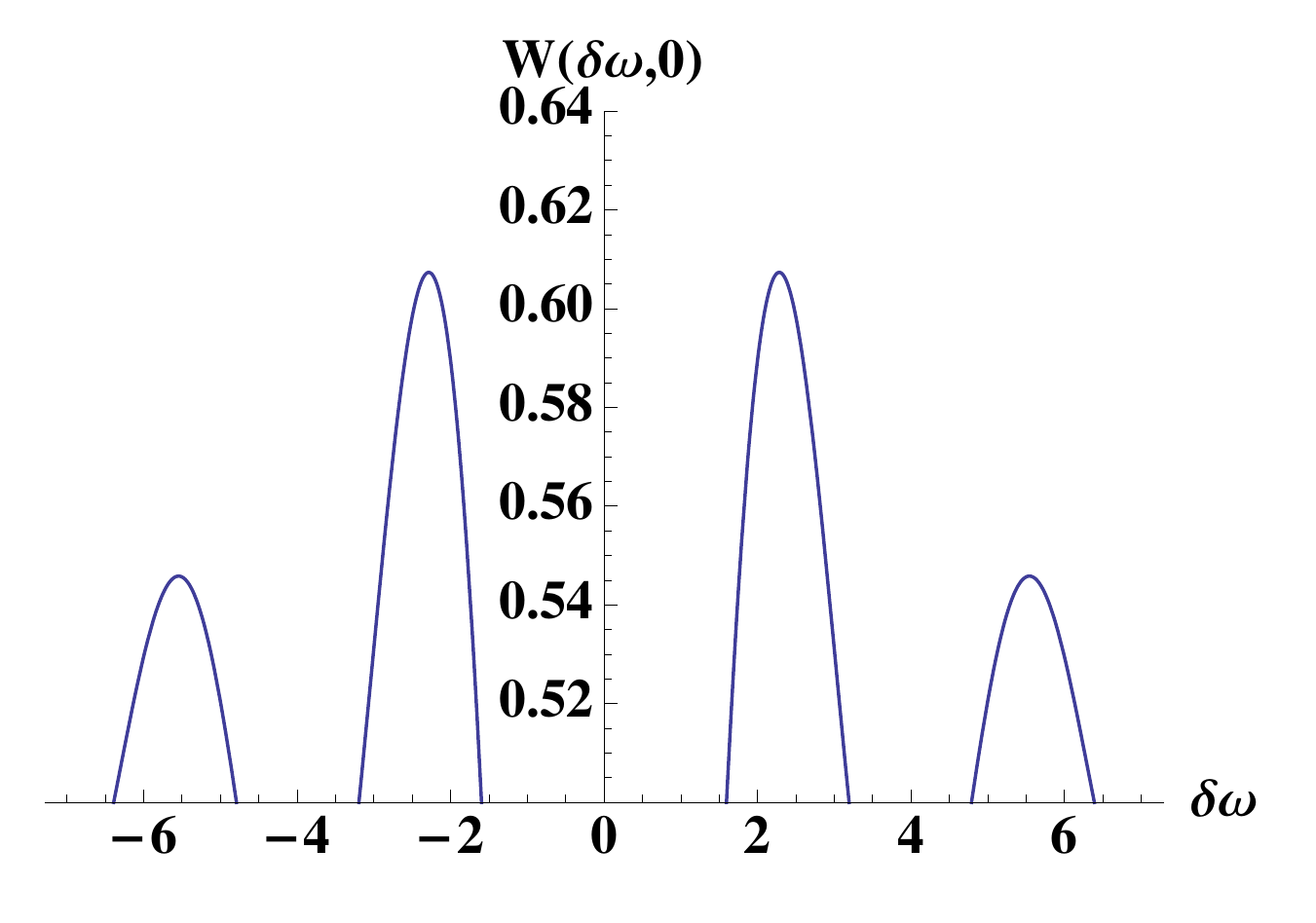} 
\end{array}$
\caption{Wigner function (normalized to $\pi$) and region  $I(\mu_i, \delta_i)>1/2$ in some chosen variable ($i=x$,$y$, $\omega$, $\tau$) for three output wave functions of the bi-photon in the SPDC process. Variables  in all the plots  are in units of the relevant physical parameters: $\mu_x \left[ \sqrt{kL^{-1}}\right ]$, $\delta_x  \left[ \sqrt{k^{-1}L}\right ]$, $\mu_{\omega}\left[ c L^{-1}(n_1-n_2)^{-1}\right ]$ and $\delta_{\tau}\left[ c^{-1} L(n_1-n_2)\right ]$ .  (a) Output of a cw pumping creating entanglement in the transverse momentum (TM) distribution. Violation of about $12\%$ can be obtained with displacements in the TM axis, as shown in (b).  (c) and (d) Schr\"odinger cat state in the TM with $\Delta p_p=5\sqrt{k/L}$. We see that a violation of over 0.9 is obtained. (e) and (f): Frequency entangled states produced in the SPDC process through pulsed pumping. \label{fig2}}
\end{figure}

We now illustrate our results by studying in more details some  examples. Let us first consider entanglement in TM naturally produced in cw SPDC using a gaussian pump. In this case, we have, in \eqref{omegay}, that $F_+({\bf p_+})=\frac{1}{\sqrt{2\pi}w_p}e^{\frac{-|{\bf p_+}|^2}{w_p^2}}$ and $F_-({\bf p_-})=\sqrt{\frac{L}{k}} {\rm sinc}{\left (\!\frac{|{\bf p_-}|^2L}{k}\!\right )}$, where $w_p$ is the width of the pumping beam momentum distribution, $k$ is the wave number of the pumping beam and $L$ is the non linear medium's length.  For simplifying reasons, we will only study coordinate $y$. This can be done by fixing $\mu_x$  and $\delta_x$ in $x$ and scanning only the $y$ phase space, i.e., varying $\mu_y$ and $\delta_y$ only.  Thus, $\pi W_{x+}(\mu_x, \delta_x)$ is a multiplicative constant.  In the studied example, since the pump is gaussian, this constant is necessarily positive, and can be set to 1 by a proper choice of $\mu_x$, $\delta_x$ and the width of the pump. Thus, Eq.~\eqref{eq:2d} directly provides $W_{y-}(\mu_y, \delta_y)$. Supposing, for simplicity,  that spatial coordinates are separable, we can apply Eq.~\eqref{omegay} to compute it, with $F_{y-}(\mu \pm p_-)=\sqrt{\frac{L}{k}}{\rm sinc}{\left (\!\frac{(\mu \pm p_-)^2L}{k}\!\right )}$. Corresponding results are shown in Fig.~(\ref{fig2}a) and (\ref{fig2}b), showing $W_{y-}(\mu_y, \delta_y)$ and $I(\mu_y, \delta_y)$. Using realistic parameters, we  can have  $I(\mu_y,\delta_y)=0.56$ for $\delta_y\approx 0.1$~mm, which fits well in the width of the transverse position distribution of the photon pairs. The relative violation of the separability threshold is of over 10~\%. Usually, this function is approximated by a Gaussian, and the relatively high violation of the entanglement witness shows the limitations of this approximation.   

Our results can be further exploited by modifying the pumping configurations and measuring the biphoton Wigner function that depends on the pump profile, \textit{i.e.},  $W_{x+}(\mu_x,\delta_x)$. We can, for instance, create  Schr\"odinger cats  \cite{SCHRO} in the TM space and directly probe their Wigner function. This can be done by coherently splitting the pumping beam in two and displacing one with respect to the other in momentum space, for instance with the help of Spatial Light Modulators (SLM). As a consequence, the TM distribution will be  centered in two different points, which distance we denote by $\Delta p_p$. The biphoton Wigner function in his case is as in Fig.~(\ref{fig2}c), where we considered $\Delta p_p= 5\sqrt{k/L}$. The entangled biphoton is highly non-Gaussian and violates the proposed witness by over 80\% (see Fig.~(\ref{fig2}d)). This configuration can also be useful  to study the decoherence of entangled non-Gaussian states through the Wigner function \cite{BUONO12}. 

Let us remark that previous results, as in  \cite{STEVE2} can now be re-interpreted with the present formulation. We see that in \cite{STEVE2}, as in the usual HOM experiment, the Wigner function at point $I(0,0;0,0)$ was measured (see \cite{SI} for a revision of these results using the present formulation). 

We provide now a second example of an experimental set-up where our results can be applied. It consists in SPDC generated from a pulsed pump in semi conductor waveguides \cite{SARA, ADELINE}. In this case, pairs of photons entangled in frequency are created, so the space of CV is one dimensional only. The output  wave function is such that $F_+(\omega_+)=\frac{1}{\sqrt{2\pi}w_p}e^{-\frac{\omega_+^2}{w_p^2}}$ and $F_-(\omega_-)=\sqrt{\frac{(n_1\!-\!n_2)L}{2\sqrt{2}c}}{\rm sinc}{\left (\!\frac{(n_1\!-\!n_2)L\omega_-}{2\sqrt{2}c}\!\right )}$ where $n_i$ are the refractive index of the medium for the $i$-th photon, $L$ is the medium's length,  $\omega_\pm=\omega_1\pm\omega_2$ where $\omega_i$ is the $i$-th photon frequency. There is a complete analogy between these functions and the one dimensional TM case, and the same reasoning can be applied leading to the measurement of the Wigner function. However, scanning  the whole phase space in this case demands using optical elements leading to  frequency displacements $\mu_{\omega}$ and frequency proportional dephasing $\delta_{\tau}$. Frequency displacements can be realized using techniques as the one demonstrated in \cite{APL}, while $\delta_{\tau}$ displacements can be done either by time-delaying one arm of the interferometer or by using linear optics elements. Expected results with the considered functions are depicted in Figs. ~(\ref{fig2}e), (\ref{fig2}f)).  In \cite{ANDREAS},  a state analogous to the Schr\"odinger cat in Fig.~(\ref{fig2}c), was created by pumping the medium with a Gaussian beam in a regime where two different phase matching conditions apply. A violation of less than $4\%$ was observable using displacements in the $\delta_\tau$ axis only by time delaying one photon(equivalent to displacements in $\delta_x$ in Figs.~(\ref{fig2}c), (\ref{fig2}d)). Negative points were observed which we can now interpret as interference fringes of the Wigner function of a biphoton Schr\"odinger cat state (see \cite{SI} for further discussion). 

In conclusion, we have provided a new interpretation of the coincidence probability in the HOM experiment in terms of a biphoton Wigner function. We adapted the HOM set-up so as all the points of the Wigner function can be detected by using available linear optics elements. Negative points of the Wigner function are associated to non-Gaussian entanglement, that can be directly detected in this type of experiment. For Gaussian entanglement, the information provided by the Wigner function can also be used applying other entanglement criteria. We analyze previous experimental results using the new perspective provided by our formulation,  that can be generalized to an arbitrary number of photons \cite{SI} or to other quantum particles with either bosonic or fermionic statistics satisfying the conditions established in the present work \cite{Beige, Electrons}. Our results open the path to realizing new fundamental tests of quantum mechanics in a simple and currently used optical setup.

{\bf Acknowledgements} The authors acknowledge A. Orieux and A. Mandilara for fruitful discussions. This work was partially financially supported by ANR/CNPq HIDE and CAPES/COFECUB 640/09. SPW and AZ thank the  FAPERJ and the INCT-Informa\c cao Qu\^antica for financial support. S.P.W. was also supported by the FET-Open Program, within the 7th Framework Programme of the European Commission under Grant No. 255914 (PHORBITECH).


\begin{thebibliography}{99}


\bibitem{HOM} C. K. Hong, Z. Y. Ou, L. Mandel, Phys. Rev. Lett. {\bf 59}, 2044, (1987).
\bibitem{WIGNER} E. P. Wigner, Phys. Rev. {\bf 40}, 749 (1932). 
\bibitem{SIMON} R. Simon, Phys. Rev. Lett. {\bf 84}, 2726 (2000); L. M. Duan, \textit{et al.}, Phys. Rev. Lett. {\bf 84}, 2722 (2000).
\bibitem{ANDREAS} A. Eckstein and C. Silberhorn, Opt. Lett. {\bf 33}, 1825 (2008). 
\bibitem{STEVE2} S. P. Walborn, \textit{et al.}, Phys. Rev. Lett {\bf 90}, 143601 (2003). 

\bibitem{Bouwmeester97} D. Bouwmeester {\it et al.}, Nature {\bf 390}, 575 (1997). 
\bibitem{Gisin02} N. Gisin, \textit{et al.}, Rev. Mod. Phys. {\bf 74}, 145 (2002). 
\bibitem{Walther05} P. Walther {\it et al. }, Nature {\bf 434}, 169 (2005). 
\bibitem{REPEATERS} K. F. Reim, \textit{et al.},
Phys. Rev. Lett. {\bf 107}, 053603 (2011).
\bibitem{BELL} A. Aspect, P. Grangier and G. Roger, Phys. Rev. Lett. {\bf 49}, 91 (1982); G. Weihs, \textit{et al.}, Phys. Rev. Lett. {\bf 21}, 5039 (1998).  
\bibitem{VOGEL} J. Sperling and W. Vogel, Phys. Rev. A {\bf 79}, 052313 (2009). 
\bibitem{TAKAHASHI09} J. Lee and H. Nha, Phys. Rev. A {\bf 87} 032307 (2013). 
\bibitem{COMP} H. Jeong, M. S. Kim and J. Lee, Phys. Rev. A {\bf 64}, 052308 (2001).
\bibitem{DALVIT} F. Toscano, \textit{et al.}, Phys. Rev. A {\bf 73}, 023803 (2006).
\bibitem{kaled}
K. Dechoum, M. D. Hahn, R. O. Vallejos, and A. Z. Khoury, Phys. Rev. A  \textbf{81}, 043834 (2010).
\bibitem{REVIEW} S. P. Walborn, C. H. Monken, S. P\'adua and P. H. Souto Ribeiro, Phys. Rep. {\bf 495}, 87 (2010). 


\bibitem{BANAZSEK} E. Mukamel, K. Banaszek, I. A. Walmsley and C. Dorrer, Opt. Lett. {\bf 28}, 1317 (2003).
\bibitem{BANAZSEK2} B. J. Smith, B. Killett, M. G. Raymer, I. A. Walmsley and K. Banaszek, Opt. Lett. {\bf 30}, 3365 (2005). 




\bibitem{STEVE}  D. Tasca, \textit{et al.}, Phys. Rev. A {\bf 83}, 052325 (2011).    

\bibitem{SI} See Supplemental Material. 
\bibitem{VANENK} M. Stobi\'nska and K. W\'odkiewicz, Phys. Rev. A {\bf 71}, 032304 (2005); M. R. Ray and S. J van Enk, Phys. Rev. A {\bf 83}, 042318 (2011). 

\bibitem{PIRES} H. Di Lorenzo Pires, H. C. B. Florijn, and M. P. van Exter, Phys. Rev. Lett. {\bf 104}, 020505 (2010).
\bibitem{COMMENT} The reflection asymmetry of the $x$ coordinate makes that Eq. (3) is strictly valid for the $y$ coordinate only.  Eq. (4) is obtained through the analogous of Eq. (3), taking into account the reflection properties of $x$. 
\bibitem{LAW} C. K. Law and J. H. Eberly, Phys. Rev. Lett. {\bf 92}, 127903 (2004). 


\bibitem{SCHRO} E. Schr\"odinger, Naturwissenschaften {\bf 23}, 807, 823, 844 (1935). 
\bibitem{BUONO12} D. Buono, \textit{et al.}, Phys. Rev. A {\bf 86}, 042308 (2012). 

\bibitem{SARA} X. Caillet {\it at al.} Opt. Express {\bf 18}, 9967-9975 (2010).
\bibitem{ADELINE} A. Orieux  {\it at al.}, Phys. Rev. Lett. {\bf 110}, 160502 (2013).
\bibitem{APL} S. Preble, \textit{et al.}, Appl. Phys. Lett. {\bf 101},  171110 (2012). 
\bibitem{Beige} Y. Liang and A. Beige, New J. of Phys. {\bf 7}, 155 (2005). 

\bibitem{Electrons} E. Bocquillon, V. Freulon, J.-M. Berroir, P. Degiovanni, B. Pla\c cais, A. Cavanna, Y. Jin, and G. F\`eve, ScienceÊ {\bf 339}, 1054Ê (2013).



\end{thebibliography}
\end{document}